\documentclass[12pt]{iopart}

\usepackage{graphicx}% Include figure files

\begin{document}
\title{Are we van der Waals ready?}
\author{T.~Bj\"orkman$^1$, A.~Gulans$^1$, A.~V.~Krasheninnikov,$^{1,2}$ and R.~M.~Nieminen$^1$}
 \address{$^1$ COMP - Aalto University School of Science, P.O. Box 11100, 00076 Aalto, Finland}
 \address{$^2$ Department of Physics, University of Helsinki, P.O. Box 43 00014 Helsinki, Finland}
 \ead{torbjorn.bjorkman@aalto.fi}
\date{\today}

\begin{abstract}
We apply a range of density-functional-theory-based methods capable of describing van der Waals interactions to weakly bonded layered solids in order to investigate their accuracy for extended systems. The methods under investigation are the local density approximation, semi-empirical force fields, non-local van der Waals density functionals and the random-phase approximation. We investigate the equilibrium geometries, elastic constants and the binding energies of a large and diverse set of compounds and arrive at conclusions about the reliability of the different methods. The study also points to some directions of further development for the non-local van der Waals density functionals.
\end{abstract}
\submitto{\JPCM}
\maketitle

\section{Introduction}
While graphene has attracted a lot of attention during the past years, the materials science community starts to move the focus to other two-dimensional materials with interesting and potentially useful properties\cite{novoselov2005,mos2,coleman1,mas-balleste2011}. Unfortunately, computational investigation of such systems within the density functional theory (DFT) is not always straightforward. A particular difficulty is the description of the van der Waals (vdW) interaction, which typically plays a crucial role in binding of two-dimensional sheets to a substrate and in surface functionalisation with organic molecules. The source of the problems are the local approximations that are conventionally applied in DFT, while the vdW interaction is an intrinsically non-local correlation effect\cite{london1937,dzyaloshinskii1961,dobson2012}. In order to overcome this issue, a number of methods\cite{llvdw1,llvdw2,vv10,grimme,Becke2005a,Tkatchenko2009,Marini2006,Furche2008} of different complexity have been developed during the past decade. These methods rely on different approximations, which have been proposed having either solid-state or molecular applications in mind. Presently, the accuracy of the approaches has been assessed almost exclusively in test cases for small molecules, largely because high-accuracy reference data can be obtained only for these systems. Based on this limited experience, it is common to anticipate qualities of a particular method in applications to extended systems. But are the existing methods equally good for solids, molecules and molecules adsorbed on surfaces of solids?

In this Paper, we consider four popular computational approaches for treating the vdW interaction, the local-density approximation (LDA)\cite{Ceperley1980,Perdew92-LDA}, Grimme's semi-empirical force-field corrections (DFT-D)~\cite{grimme}, non-local van der Waals density functionals (vdW-DF)~\cite{llvdw1,llvdw2,vv10} and the random-phase approximation (RPA)~\cite{pines-nozieres_book,langreth-perdew1977}, by applying them to 74 layered solids. As the primary objectives of the study, we select the equilibrium interlayer separations and layer thicknesses, the interlayer binding energies and the $C_{33}$ elastic constants. As in any solid, the equilibrium geometry is a critical quantity in layered solids, since features of the electronic structure, such as the band gap, may critically depend on the interlayer separation\cite{li2007,kadantsev2012}. The interlayer binding energy is a key quantity for the study of exfoliation of two-dimensional compounds\cite{coleman1}. The calculated equilibrium geometries are assessed by comparison with experimental data, but since experimental interlayer binding energies are not available for any material except for graphite\cite{girifalco1956,benedict1998,zacharia2004}, we perform random-phase approximation (RPA) calculations for a subset of 26 solids. While RPA is an approximate method and its performance also has to be comprehensively tested, recent successes\cite{harl2008,seb,rpa3,Eshuis2011} encourage us to believe that, for non-covalent interactions, RPA is the most systematic and accurate one among the methods considered here. Our RPA calculations are the first application of this method to a wide range of solids covering a significant part of the Periodic Table. 

The paper is divided into a Methods section, where we describe the selection of the investigated compounds, details of the computational procedure and a brief introduction to the most important technicalities of the methods. For a thorough review of RPA and vdW-DF we refer the reader to a recent paper by Dobson and Gould\cite{dobson2012}. The Results section presents figures and tables of the evaluation data, organized by the investigated property. Finally, the Discussion section evaluates the performance of the different methods, in particular discussing the problems associated with LDA in vdW systems, the parametrization of DFT-D for solids and the problems associated with selecting an appropriate parent functional of the non-local vdW functionals to account for the exchange.

\section{\label{sec:methods}Methods}
\subsection{Test systems}
We compare results for 74 different compounds identified in the Inorganic Crystal Structure Database (ICSD)\cite{icsd} by data filtering of layered three dimensional structures with interlayer bond lengths that indicate that they are likely to be dominated by non-covalent interactions. The acquired systems are very diverse in character, containing anything from magnetic metals to wide-gap insulators, single atomic layers, such as graphite or BN, as well as thicker layers, In$_2$Zn$_2$S$_5$ being the thickest with a layer thickness of 12.4~\AA. For computational reasons, we selected only high symmetry structures corresponding to systems consisting of either hexagonal or quadratic planes and a small number of the systems found needed to discarded on computational grounds as the time required for the calculations were too long. This automated selection procedure ensures against our own biases when selecting the compounds to be used in testing different methods, and enables us to make general statements about the properties of the different methods. The complete list of compounds is given in Table~\ref{allthecompounds} and tabulated values for all calculated quantities are listed in the Supplemental Material of Reference~\cite{bjorkman2012a} where also a more detailed description of the selection procedure can be found.

\subsection{Brief description of the methods}

It has been known for a long time that, despite lack of formal justification, the local-density approximation (LDA) provides a reasonable description of the bond lengths and binding energies for many vdW-bonded layered systems, such as graphite\cite{yin1984,jansen1987}. At the same time, it has been recognized that this is just a fortuitous coincidence\cite{harris1985}, since the binding in the LDA picture stems from the exchange, while the vdW interaction is a correlation effect\cite{harris1985,gerber2007,murray2009}. Despite this knowledge, numerous studies have applied LDA to emulate the vdW interactions in layered structures. In light of this, we include this method in our benchmark calculations and evaluate how good or bad LDA is for layered materials less well known than graphite to get further knowledge about how common such fortuitous coincidences are.

The semi-empirical approach by Grimme known as DFT-D\cite{grimme} is a straightforward force-field correction based on the assumption that the total dispersion interaction between larger molecules or solids can be described as a sum of contributions from all pairs of atoms.  Each pair contributes a term proportional to the inverse sixth power of its interatomic distance, $R$,
\begin{equation}
\label{eq:dft-d}
E_{vdW} = -s_6 \sum_{pairs} \frac{C_6}{R^6} \, f_{dmp}(R).
\end{equation}
The formula also contains an empirical overall scale factor, $s_6$, which is different for each exchange-correlation functional, the atomic $C_6$ coefficients and a damping function, $f_{dmp}$, that prevents (\ref{eq:dft-d}) from diverging at small $R$.

A different strategy is pursued in the construction of the vdW density functionals (vdW-DF\cite{llvdw1}, vdW-DF2\cite{llvdw2} and VV10\cite{vv10}). These methods obtain the vdW interaction from the electron density $\rho(r)$ via the genuinely non-local correlation functional
\begin{equation}
E^\mathrm{c}_\mathrm{nl} = \int\int dr dr' \rho(r)\Phi^c(r,r')\rho(r'),
\end{equation}
where $\Phi^c(r,r')$ is a kernel function derived from a local polarizability model\cite{vydrov2010a} using a number of approximations\cite{llvdw1,llvdw2,vv10,dobson2012}. 
By construction, $E^\mathrm{c}_\mathrm{nl}$ vanishes for a uniform electron density, hence, the correlation energy is complemented by the LDA contribution. Since $E^\mathrm{c}_\mathrm{nl}$ contains the necessary ingredients for vdW forces, an additional attraction stemming from the exchange functional as in LDA is undesirable. For this reason, the original choice for the exchange to be used with $E_{vdW}$ was the revPBE functional\cite{zhang1998}, which is almost free from any spurious binding. Soon it was realised that revPBE is typically too repulsive in the vdW regime\cite{andrisvdw,murray2009}, and a large number of other options for the exchange part of the functional have been proposed, including a revised version of the PW86 functional, PW86R\cite{murray2009,llvdw2,vv10}, PBE\cite{andrisvdw}, optimized versions of PBE\cite{Klimes2010} and long-range corrected hybrid functionals\cite{sato2007,vv10}. In this Paper, we explore the performance of vdW-DF combined with RPBE\footnote{RPBE by Hammer et al.\cite{hammer1999} is a functional constructed to mimic features of revPBE while obeying the Lieb-Oxford criterion\cite{lieb-oxford}.} and PBE, vdW-DF2 with PW86R and VV10 with PW86R, thus allowing for direct comparisons of the different components of some of the functionals (see the discussion in Section \ref{vdw_df_discussion}).

The adiabatic-connection fluctuation-dissipation theorem (ACFDT)\cite{langreth-perdew1977} is a powerful technique that, in principle, allows to obtain the exact exchange-correlation energy within the DFT framework.  The method uses the standard integration over the coupling constant $\lambda$\cite{pines-nozieres_book} to construct the interacting system from the non-interacting one, here taken to be the Kohn-Sham system. After inserting the exact exchange in ACFDT, the exact correlation energy then can be expressed as
\begin{equation}\label{rpacorr}
E_c = -\mathrm{Tr}\int_0^{\infty} \frac{d\omega}{2\pi} \int_0^1\frac{d\lambda}{\lambda} \left(\chi_{\lambda}(i\omega)-\chi_{KS}(i\omega)\right)V_\lambda,
\end{equation}
where $V_\lambda$ is $\lambda$ times the Coulomb potential and $\chi_{\lambda}$ and $\chi_{KS}$ are the frequency-dependent density-response functions for the interacting and Kohn-Sham systems, respectively. If the so-called exchange-correlation kernel\cite{dobson-wang2000}, $f_{xc}^{\lambda}$, is known, we can obtain $\chi_{\lambda}$ from Dyson's equation
\begin{equation}\label{response}
\chi_{\lambda} = \chi_{KS} + \chi_{KS}(V_\lambda + f_{xc}^{\lambda})\chi_{\lambda},
\end{equation}
but in practice, (\ref{rpacorr}) is untractable for real systems, unless $f_{xc}^{\lambda}$ is approximated. Presently, we consider RPA (also sometimes called direct-RPA), where the exchange-correlation kernel in (\ref{response}) is neglected altogether. This approximation makes it possible to integrate the coupling constant analytically and simplifies numerical efforts. 
Nevertheless, RPA calculations present a formidable task, which is currently a massive obstacle for applying RPA for a wider circle of applications.

\subsection{Computational procedure}\label{computational_procedure}

Calculations were performed using the projector augmented wave method as implemented in the electronic structure package VASP\cite{vasp3,kresse1999}, with an in-house implementation of the vdW-DF method\cite{andrisvdw}. Crystal geometries were automatically generated from database searches using the program CIF2Cell\cite{cif2cell}. The projector-augmented wave (PAW) potentials  from the library distributed with the VASP code\cite{kresse1999} were used and plane wave cutoffs were initially selected as 1.5 times the default cutoff, subsequently increased in individual cases if there were apparent convergence problems. The convergence was more carefully tested for a small subset of compounds. Compounds containing elements in the $3d$ series from Cr to Ni were calculated in the ferromagnetic mode. Brillouin zone integrations were performed using Gaussian smearing with a smearing width of 0.1eV, using a uniform mesh with the number of points selected to give a distance of 0.2\AA$^{-1}$ between the mesh points for non-magnetic calculations and 0.15\AA$^{-1}$ for magnetic calculations. 

Due to the overwhelming computational expenses, the RPA calculations were carried out using different computational settings.
The settings were 420 eV and 0.2--0.3\AA$^{-1}$ for the plane-wave energy cut-off and the $\mathbf{k}$-point spacing, respectively. The corresponding exact-exchange calculations were performed using the same plane-wave cut-off, but the $\mathbf{k}$-point spacing was further refined for semiconductors or left the same for metals, as it was described in Reference~\cite{rpa1}.
The reference Kohn-Sham states were obtained using the PBE exchange-correlation functional.
In the VASP code, the density-response functions are constructed using the relation by Adler and Wiser\cite{Adler1962,Wiser1963} and they take form $\chi_{KS}^\mathbf{k}(\mathbf{G},\mathbf{G^\prime},i\omega)$, where $\mathbf{G}$ and $\mathbf{G^\prime}$ are multiples of the reciprocal lattice vectors and $\mathbf{k}$ is a point within the Brillouin zone.
This representation, in principle, requires an infinite number of plane waves, but, in practice, their number is restricted by the energy cut-off $E_\mathrm{cut}$ so that all vectors $\frac{\mathbf{G}^2}{2}>E_\mathrm{cut}$ are discarded.
Equivalently, one can use the maximum wavenumber $q$, then, the condition above translates into $|\mathbf{G}|>q$.
It was shown that these basis truncation parameters have a strong influence on the correlation energy. In particular, Harl and Kresse\cite{harl2008} have suggested that the correlation energy converges as
\begin{equation}
\label{convergence1}
E_\mathrm{c}^\mathrm{RPA}(q)=E_\mathrm{c}^\mathrm{RPA}(q=\infty)+A/q^3,
\end{equation}
where A is a constant and $q$ is the cut-off wavenumber that can be related to the cut-off energy through the relation $E_\mathrm{cut}=q^2/2$. However, it can be shown\cite{rpa1} that (\ref{convergence1}) can be extended to 
\begin{equation}
\label{convergence2}
E_\mathrm{c}^\mathrm{RPA}(q)=E_\mathrm{c}^\mathrm{RPA}(q=\infty)+A/q^3+B/q^5+C/q^6+\cdots,
\end{equation}
where A, B and C are constants. 
Typically, energy differences have better convergence properties than the total energies themselves, and when Equation (\ref{convergence2}) is investigated for the energy differences of the interlayer binding energies, we find numerically that the terms containing $q^{-3}$ and $q^{-6}$ vanish. This empirical observation allows us to write the following relation for the energy differences,
\begin{equation}
\label{convergence3}
\Delta E_\mathrm{c}^\mathrm{RPA}(q)\approx \Delta E_\mathrm{c}^\mathrm{RPA}(q=\infty)+\alpha/q^5+\beta/q^7+\cdots,
\end{equation}
where $\alpha$ and $\beta$ are constants. In practical calculations, we have calculated RPA correlation energies using different cut-off energies and have used them for fitting of (\ref{convergence3}). This procedure allowed us to obtain accurate estimates of the complete-basis limit with cut-off energies as low as 100--150~eV, which are significantly lower than those previously used in References \cite{harl2008,rpa1}. Translated into computational effort, this procedure allows us to obtain the binding energies cheaper by an order of magnitude without sacrificing the accuracy. 

The interlayer binding is investigated by varying the $c$ axis length and calculating the corresponding total energies. The intralayer coordinates were allowed to relax at each $c$ axis length, but the in-plane lattice constant was kept fixed at its  experimental value. The in-plane lattice constant is dominated by covalent bonding, for which the errors are very much smaller than the errors from the treatment of the vdW interactions, and tests for the transition metal dichalcogenides showed that allowing for full relaxation has very small impact on the results presented here. However, this simplification will induce additional uncertainty in the calculated thicknesses of the layers, since errors in the bond lengths that would result in a different in-plane lattice constant can be compensated for by relaxation in the  $c$-direction. Thus we expect that the variations in intralayer thicknesses between the functionals are somewhat exaggerated in the present study.

\begin{figure}[htbp] %  figure placement: here, top, bottom, or page
   \centering
   \includegraphics[width=0.5\textwidth]{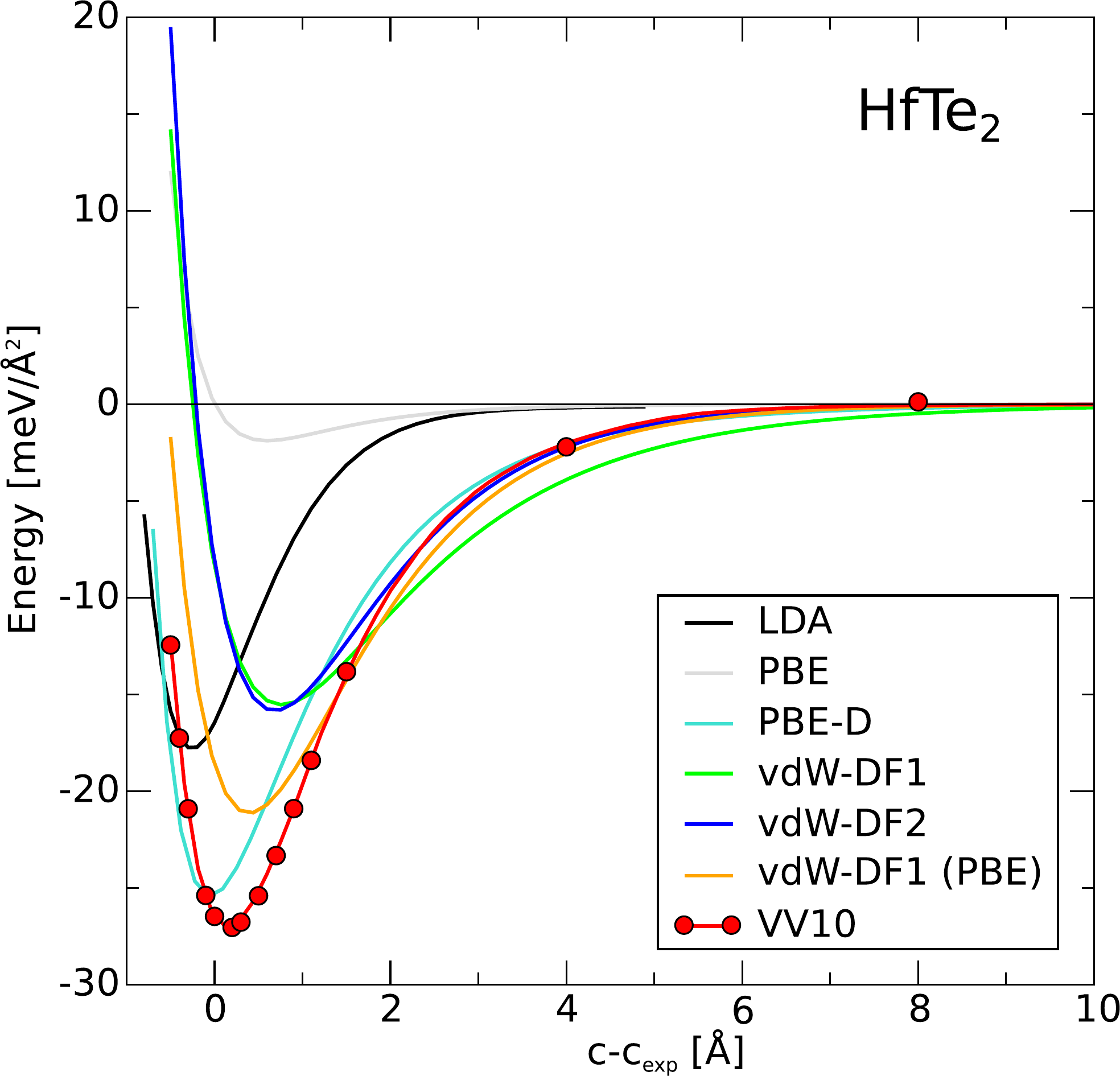} 
   \caption{Calculated energy per surface unit area of HfTe$_2$ as function of the deviation from the experimental $c$ lattice constant. For readability, only interpolated curves are shown for all functionals except VV10, where the calculated points are also shown. }
   \label{bindencurves}
\end{figure}

Figure~\ref{bindencurves} shows a typical result of the procedure, for HfTe$_2$. On the compression side there is Pauli repulsion as the densities overlap, making an exponentially rising ''exchange wall'', and on the expansion side there is a lowly decaying tail, an attractive ''van der Waals slope''. Two features need to be resolved and extracted from the data, the energy minimum and the asymptotic behaviour at large separations.  The seven data points closest to the minimum were fitted to a fourth-order polynomial, which was used to obtain the equilibrium lattice constant and the $C_{33}$ elastic constant. As the structure is stretched along the $c$ axis, the energy per atom approaches the value in an isolated layer. We determine this value by simply increasing the lattice constant until the change in energy is sufficiently small, typically at interlayer separations 10-15\AA{} larger than the experimental equilibrium distance. The popular method of fitting the large-separation tail to some known function and extracting the asymptote was tried and discarded, since it was found to induce large uncertainties. In fact, even extracting the correct power-law of the asymptote is a non-trivial task\cite{seb}.

In the RPA calculations, only a few energy points near the minimum and one point for a single layer were calculated. The intralayer geometry was kept fixed at the thickness as determined from a PBE calculation for a single layer. The binding energies are quite insensitive to this approximation, since the intralayer forces induced by the vdW interaction are zero at the minimum of the binding energy curve, and the intralayer equilibrium geometry will therefore be the same as for a single layer. However, elastic constants will be overestimated due to the stiffness of the layers. For MoS$_2$, we found the $C_{33}$ elastic constant to be overestimated by approximately 10\%, and we believe this to be the typical error.

\section{\label{sec:results} Results}

The results of the calculations are divided into the relaxed equilibrium geometries, interlayer binding energies and $C_{33}$ elastic constants. The equilibrium geometries are straightforward to compare with the experimental data from the ICSD, while for the binding energies the RPA calculations are used as benchmark. The benchmarking of the $C_{33}$ elastic constants are somewhat troublesome since, as explained above, the RPA calculations do not yield a good benchmark due to the neglect of intralayer relaxations. Experimental data for the elastic constants is unfortunately scarce and come with large uncertainties, but some general conclusions may still be drawn.

\subsection{Geometries}

\begin{figure}[htbp] %  figure placement: here, top, bottom, or page
   \centering
   \includegraphics[width=\textwidth]{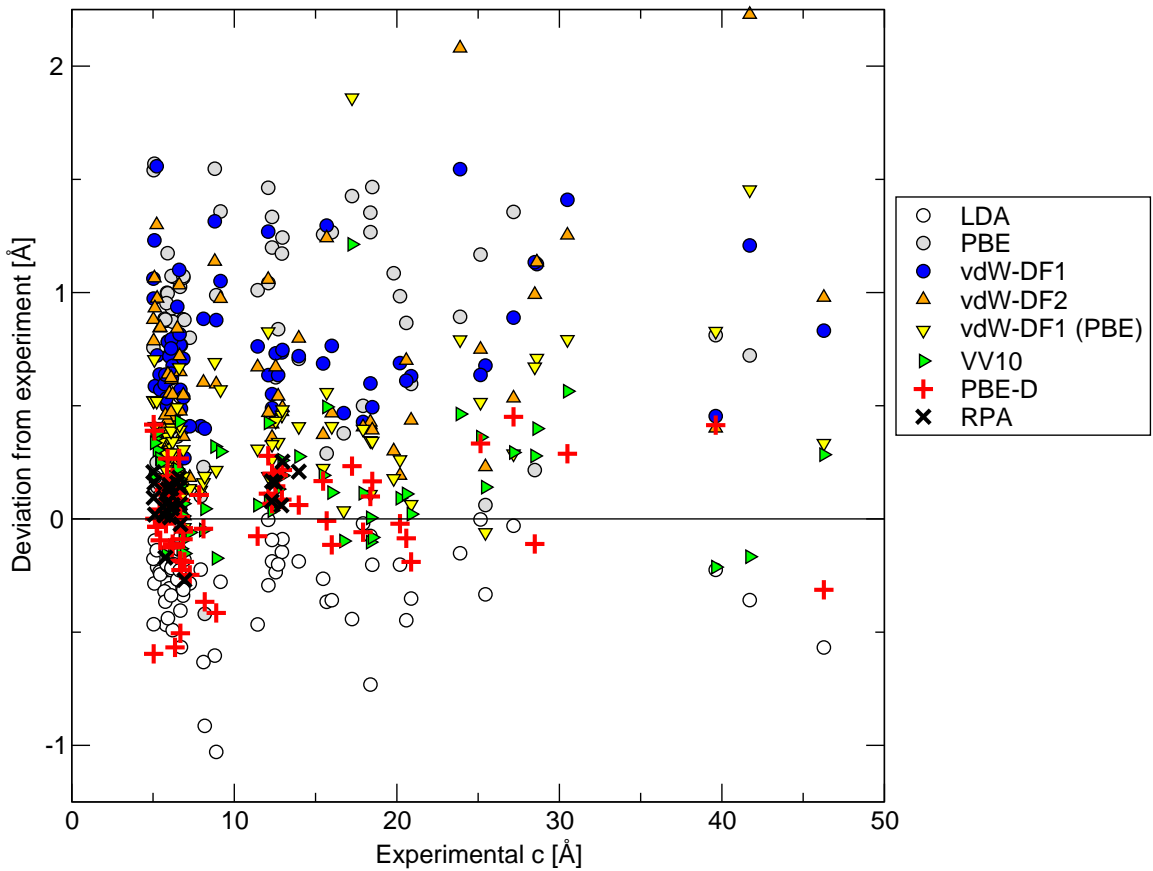} 
   \caption{Comparison of $c$ axis lengths with experimental values for the investigated functionals.}
   \label{exp-calc_c}
\end{figure}

\begin{figure}[htbp] %  figure placement: here, top, bottom, or page
   \centering
   \includegraphics[width=0.75\textwidth]{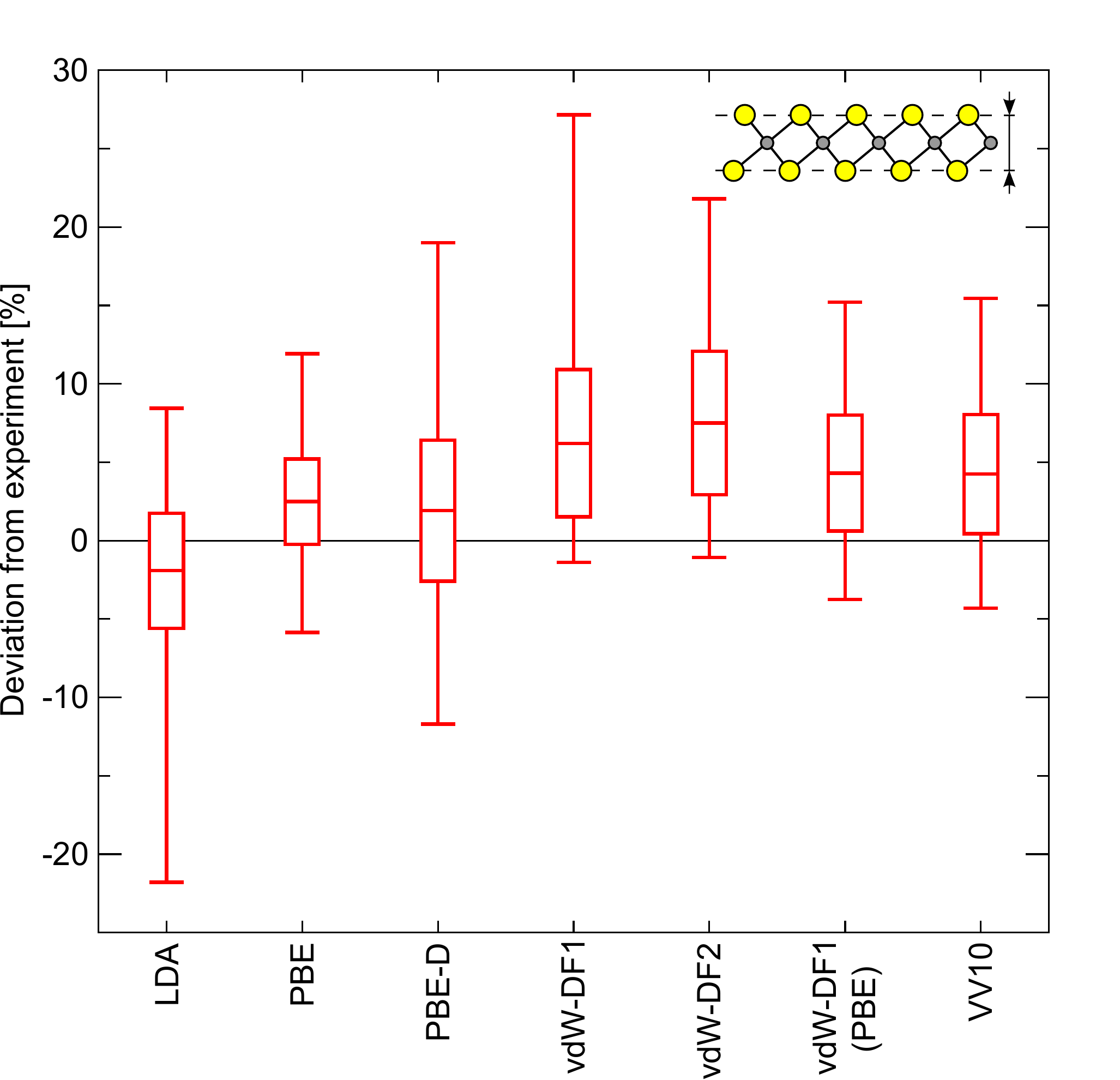} 
   \caption{Comparison of the deviation of intralayer thickness from experimentally reported values for the investigated functionals. The interlayer thickness was not relaxed in the RPA calculations, but taken from PBE calculations.}
   \label{thickdevs}
\end{figure}

\begin{figure}[htbp] %  figure placement: here, top, bottom, or page
   \centering
   \includegraphics[width=0.75\textwidth]{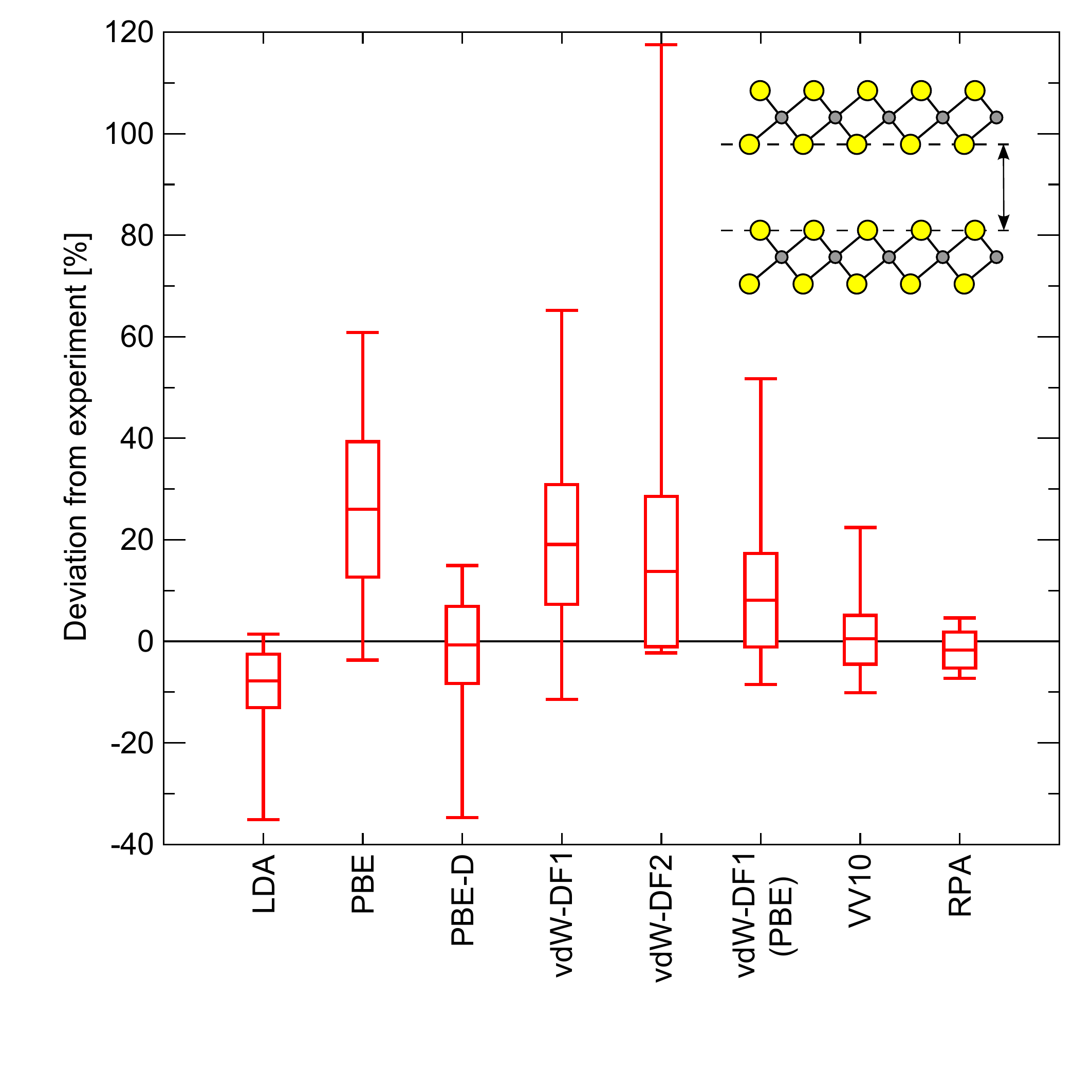} 
   \caption{Comparison of the deviation of interlayer distances from experimentally reported values for the different functionals. The RPA statistics is based on only 26 compounds.}
   \label{gapdevs}
\end{figure}

\begin{figure}[htbp] %  figure placement: here, top, bottom, or page
   \centering
   \includegraphics[width=0.75\textwidth]{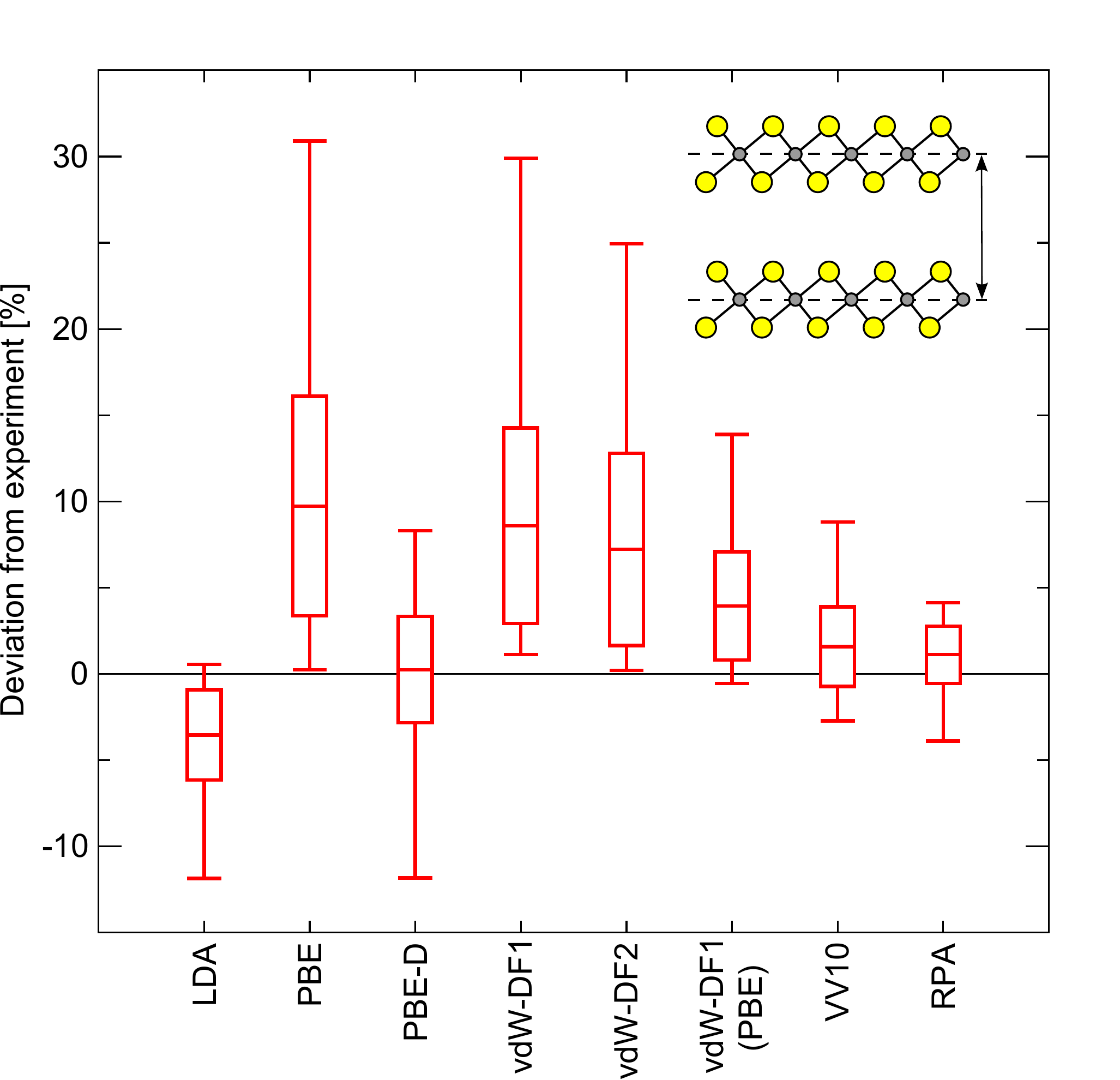} 
   \caption{Comparison of the deviation of $c$ axis lengths from experimentally reported values for different functionals. The RPA statistics is based on only 26 compounds.}
   \label{cdevs}
\end{figure}

The relaxed geometries are analysed in terms of the crystallographic $c$ axis length and its two components, the interlayer distance and the intralayer thickness. Figure~\ref{exp-calc_c} shows the results of a comparison of the experimental $c$ axis lengths to the calculated values for all different functionals in terms of the deviations from the experimentally reported value. Figures \ref{thickdevs}-\ref{cdevs} summarise the relative deviations from experiments showing boxes centred at the average deviation with a total height of two standard deviations of the distribution, and with the maximal deviations indicated by extension lines. It should be noted that the deviations of the $c$ axis lengths have a total span of more than 3\AA, or 40\%, enormous errors by the current standards of high-accuracy testing of density functionals of solids\cite{haas2009,csonka2009}. In these circumstances, the neglect of relaxation of the in-plane lattice constant and zero-point motion as well as the arbitrary choice of comparison with the most recently published experimental value are of little consequence. If Figure~\ref{exp-calc_c} is done making the comparison with the best or the worst fitting experimental number for each compound, or complete relaxation is done (this was tested for all the transition metal dichalcogenides), the effect is too small to be discernible.

Turning first to the intralayer thicknesses,  determined primarily by covalent bonding, shown in Figure~\ref{thickdevs}, we note that the intralayer thicknesses reflect the usual LDA overbinding of the covalent bonds as well as the PBE underbinding\cite{haas2009}, albeit somewhat larger than usual for the reasons discussed in Section \ref{computational_procedure}. LDA predicts intralayer thicknesses that are on average 2-3\% too small and PBE is too large by a similar amount. The average deviation of PBE-D is similar to PBE, but with a larger spread of the values. The original vdW-DF1 and the later vdW-DF2 perform significantly worse for the intralayer geometry, and while vdW-DF1 (PBE) and VV10 show improvement, they still fall short of the performance of the plain PBE functional. For the RPA, relaxation of the layer thickness was not done and instead the equilibrium geometry of a single layer from a PBE calculation was used. These results are similar to the findings of  Klime\v{s} et al.\cite{klimes2011} and Wellendorf and Bligaard\cite{wellendorf2011} for bulk solids, although the trends are more pronounced in the present inhomogenous geometry.

For the vdW-dominated interlayer separations, shown in Figure~\ref{gapdevs}, the first important thing to note is that the scale of the deviations has increased drastically. LDA is even more overbinding than at covalent bond distances, and PBE is very much underbinding, on average overestimating the interlayer distance by more than 20\%. PBE-D gives on average very good interlayer spacings, but is significantly overbinding for some compounds, as shown by the large maximal deviation. For the vdW-DF's, we can se a gradual improvement in the sequence vdW-DF1, vdW-DF2, vdW-DF1 (PBE) and VV10. The vdW-DF1 functional is in fact only a very small improvement over PBE for the interlayer separations, and vdW-DF2 improves on this only a little further (excepting the compound PbBi$_4$Te$_7$, where vdW-DF2 overestimates the interlayer separation by as much as 118\%). Using PBE exchange, we then see a significant improvement of the bond lengths for the vdW-DF1 (PBE) functional. The VV10 functional is very much superior to all the other single-particle based theories, with a narrow distribution around the experimentally reported values and only moderate maximal deviations. However, the RPA distribution is very sharp, with an average just below the experimentally reported values, as is to be expected since the experiments are generally conducted at higher temperatures. 

In Figure~\ref{cdevs} we show the results for the $c$-axis lengths. Summarizing the findings for the equilibrium geometries, the PBE functional will give much too large vdW bond lengths, while LDA is overbinding, with bond lengths that in some cases are very much too small. The PBE-D functional on average produces good equilibrium geometries, although with a tendency to sometimes produce bad results for no apparent reason. Among the vdW-DF type functionals, the original vdW-DF1 is the least accurate for all geometrical properties (disregarding outliers in the vdW-DF2 distribution of interlayer distances in Figure~\ref{gapdevs}), and the more recent VV10 gives drastically better results. For the vdW component, as measured by the intralayer gap, the accuracy of VV10 even approaches the RPA results. The vdW-DF2 shows improvement over vdW-DF1, but the vdW bond lengths are still too large. The bond lengths for vdW-DF1 can be radically improved by using PBE exchange rather than RPBE.

\subsection{Binding energies}

\begin{figure}[htbp] %  figure placement: here, top, bottom, or page
   \centering
   \includegraphics[width=\textwidth]{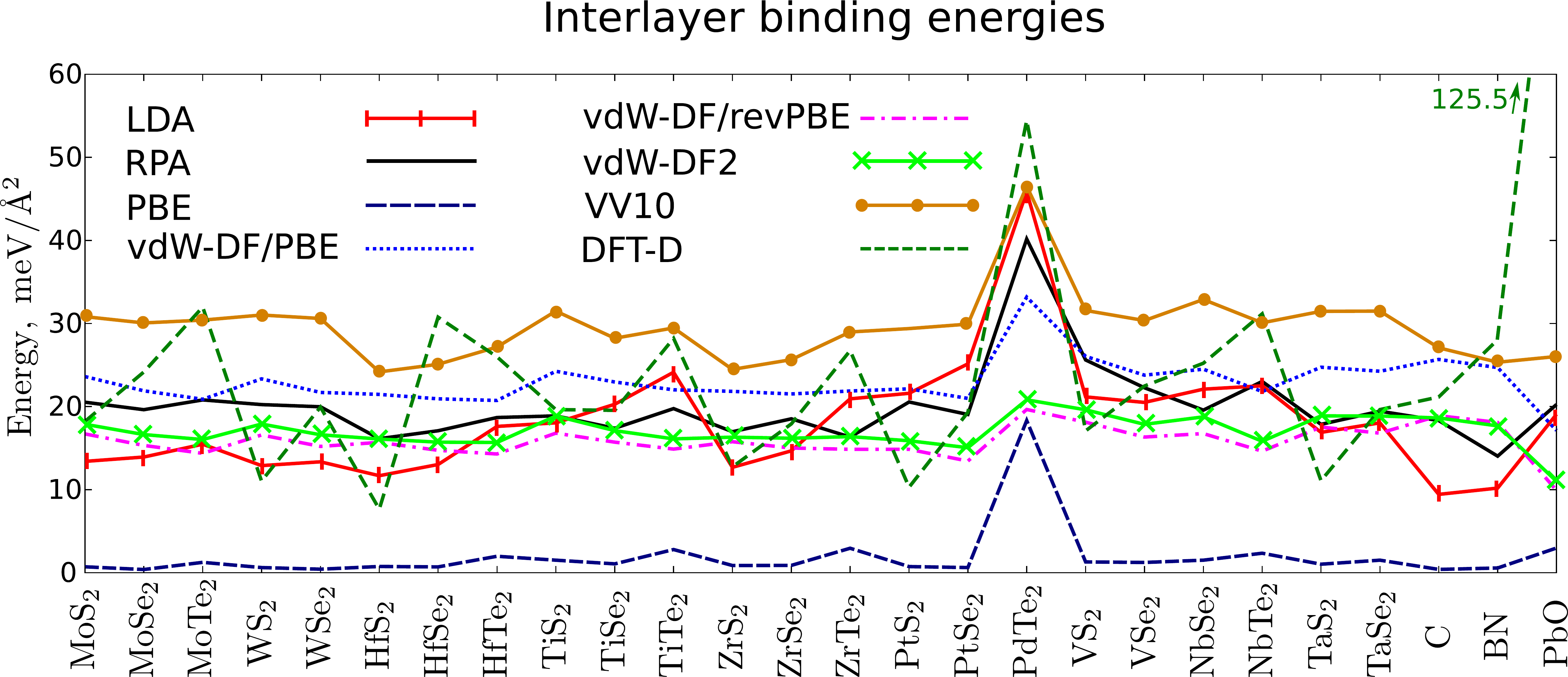} 
   \caption{The binding energies of the 26 compounds calculated with RPA, showing the comparison with the different functionals. The value for PbO for the PBE-D method that is outside the diagram is 125.5 meV/\AA$^2$.}
   \label{bindencomp}
\end{figure}

As mentioned in the Introduction, we aim to study the binding energies of the different methods using RPA as a benchmark. Figure~\ref{bindencomp} shows the outcome for 26 compounds. The RPA binding energies are quite consistently found in a range about 15-20 meV/\AA$^2$, with an outstanding exception in PdTe$_2$. As PdTe$_2$ has a significant binding energy even with the PBE functional, which otherwise gives near zero binding energies, we conclude that this is due to some weak covalent bonding occurring in this material. LDA gives binding energies  in roughly the correct interval, but does not follow the trends of RPA particularly well and there appears to be no way of telling whether binding energies will be higher or lower. PBE-D deviates rather strongly from RPA and fails both in reproducing trends and in one case, PbO, even fails to give the right order of magnitude.
vdW-DF1 and vdW-DF2 are somewhat too low, vdW-DF1 (PBE) is somewhat too high and VV10 overshoots considerably, being quite consistently 50\% too high. The vdW-DF's follow the RPA trends very closely with a few exceptions, which on closer inspection are found to correlate with particularly bad overestimations of the interlayer distances.

When comparing the distributions of the interlayer binding energies we see that RPA and vdW-DF1, vdW-DF2 and vdW-DF1 (PBE) show dense peaks, VV10 has a main peak, but with a tail at lower binding energies. LDA has a two peaks, the lower being somewhat lower than the RPA peak. PBE-D shows the most scattered distribution, with many isolated points at very high binding energies. For the vdW-DF's, we note the same trends as for the bond lengths with vdW-DF1 having the lowest binding energies and vdW-DF2, vdW-DF1 (PBE) and VV10 following in order of binding strength, with vdW-DF2 having a distribution closest to that of RPA.

\begin{figure}[htbp] %  figure placement: here, top, bottom, or page
   \centering
   \includegraphics[width=0.5\textwidth]{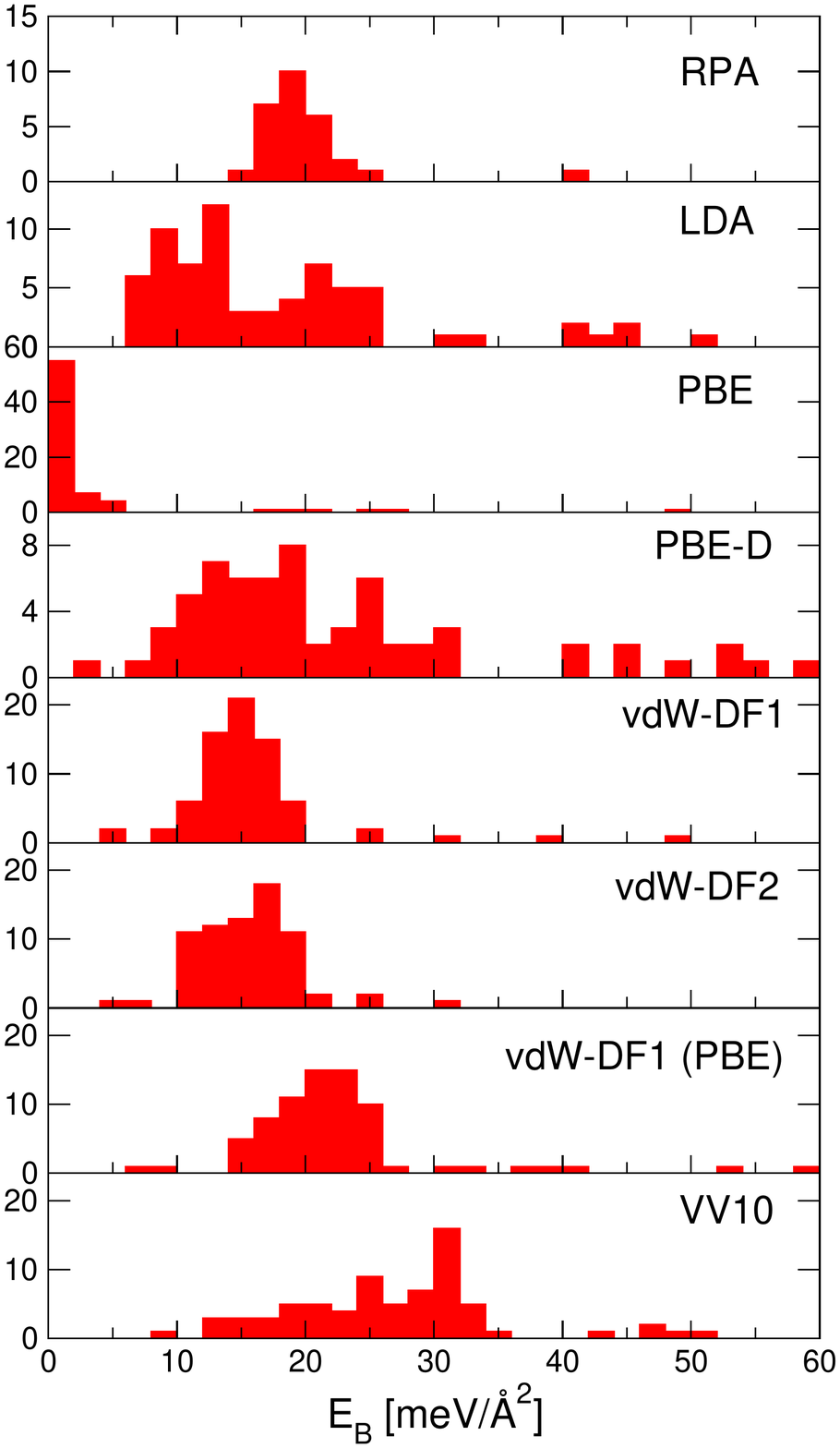} 
   \caption{Comparison of the distributions of binding energies for the different functionals. }
   \label{bindning_energy_distributions}
\end{figure}

\subsection{$C_{33}$ elastic constants}

We attempt to compare $C_{33}$ constants with experimentally reported values, but it is obvious that the experimental situation is not all that clear, and the reported experimental values often scatter significantly. This unfortunate circumstance obviously hampers any benchmarking effort, but some limited conclusions can still be drawn about the performance of the various functionals.

\begin{figure}[htbp] %  figure placement: here, top, bottom, or page
   \centering
   \includegraphics[width=0.5\textwidth]{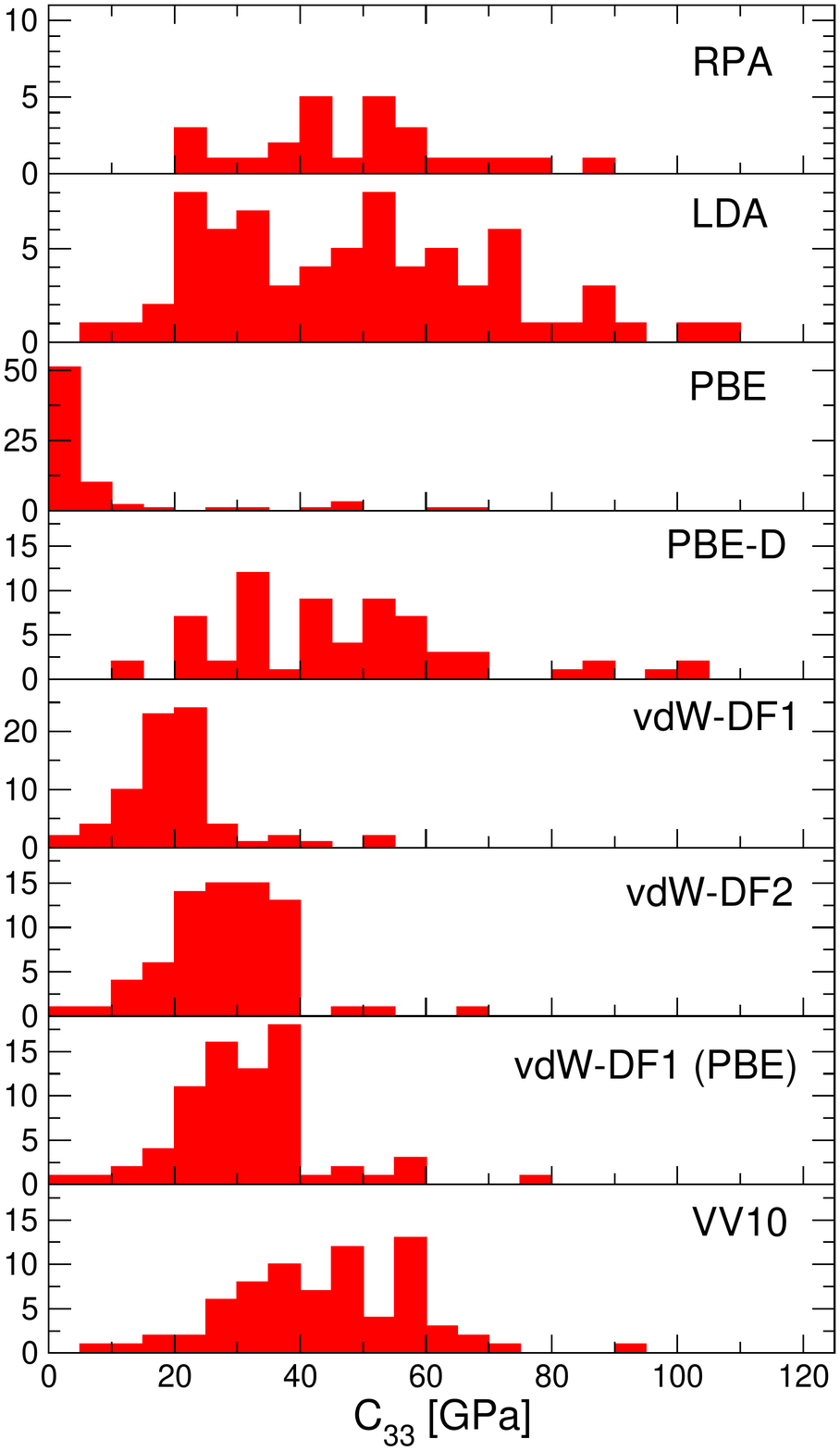} 
   \caption{Comparison of the distributions of $C_{33}$ elastic constants for the different functionals.}
   \label{c33fig}
\end{figure}

The distributions of $C_{33}$ elastic constants are shown in Figure~\ref{c33fig}. The LDA, PBE-D and RPA data span a rather large range, up to 100 GPa, while the vdW-DF's have less dispersed distributions. Just as for the binding energies, the failure of PBE to include the vdW interaction is clearly seen in the distribution of the elastic constants, which are almost exclusively in the range $< 10$ GPa. For the vdW density functionals, the elastic constants again show similar trends to the geometries and binding energies, with the functionals being progressively stiffer elastic constants in the order: vdW-DF1, vdW-DF2, vdW-DF1 (PBE), VV10.

The $C_{33}$ elastic constants for 16 of the investigated compounds for all computational methods as well as experimental data are given in Table \ref{c33}. Overall, all methods except for PBE (not shown) produce values in the correct range, but while LDA, RPA and the vdW-DF's to a reasonable degree follow the trends of the experimental data, PBE-D is doing less well. As previously mentioned, the comparison with experiment is somewhat uncertain, but it appears safe to conclude that vdW-DF1 systematically gives a too soft and VV10 a likewise too hard $C_{33}$ constant. 

\begin{table}[htdp]
\caption{$C_{33}$ elastic constants in GPa for a set of layered compounds.}
\begin{center}
\footnotesize
\begin{tabular}{c|ccccccc|c}
Compound & LDA & PBE-D & DF1 & DF2 & DF1(PBE) & VV10 & RPA & Experiment\\\hline
MoS$_2$  &  53 &   51  &  24 &  39 &     39    &   61 & 59  & 52\cite{feldman1976} \\
NbSe$_2$ &  74 &   51  &  22 &  37 &     35    &   55 & 71  & 43--60\cite{feldman1976}, 42\cite{moncton1977}, 67\cite{sezerman1980}, 50.9, 52\cite{lb} \\
TiS$_2$  &  64 &   41  &  23 &  36 &     36    &   51 & 51  & 54.6$\pm5.3$\cite{scharli1986} \\
TiSe$_2$ &  69 &   43  &  19 &  32 &     30    &   44 & 41  & 39.0$\pm3.0$\cite{scharli1986}, 39\cite{abdullaev2006}, 45.6\cite{galliardt1981}\\
HfS$_2$ & 40 & 22 & 22 & 33 & 34 & 46 & 40 & 39.4\cite{lb} \\
TaS$_2$ & 51 & 34 & 25 & 40 & 40 & 51 & 59 & 50.5$\pm$3\cite{ziebeck1977}\\
TaSe$_2$ &  53 &   76  &  22 &  35 &     37    &   31 & 80  & 54\cite{moncton1977} \\
WS$_2$ & 51 & 34 & 24 & 39 & 40 & 62 & 56 & 60$\pm$5\cite{sourisseau1991}\\
Graphite &  30 &   44  &  23 &  33 &     36    &   46 & 36  & 40.7\cite{gauster1974}, 36.5\cite{gauster1974,blakslee1970}, 38.7\cite{bosak2007}, 37\cite{wada1980,abdullaev2006}\\
BN       &  29 &   69  &  20 &  30 &     32    &   41 & 25  & 32.4$\pm$3\cite{green1976}, 35.6\cite{kelly1977},18.7\cite{duclaux1992}\\
GaS      &  37 &   31  &  24 &  34 &     35    &   46 & --  & 36\cite{abdullaev2006}, 38.5\cite{gatulle1983} \\
GaSe     &  37 &   52  &  20 &  27 &     29    &   40 & --  & 34\cite{abdullaev2006}, 35.7\cite{gatulle1983}, 31.9\cite{panella1999} \\
CdI$_2$  &  24 &   33  &  15 &  21 &     22    &   28 & --  & 22.5\cite{lb} \\
HgI$_2$  &  21 &   26  &  11 &  15 &     17    &   23 & --  & 15.3, 16.3\cite{lb} \\
SnS$_2$ & 30 & 30 & 19 & 27 & 28 & 37 & -- & 27.2\cite{lb} \\
Bi$_2$Te$_3$ & 55 & 48 & 15 & 23 & 24 & 40 & -- & 47.7\cite{lb}
\end{tabular}
\end{center}
\label{c33}
\end{table}%

\section{Discussion}

\subsection{RPA}
The present study treats RPA as a benchmark method for binding energies while at the same time evaluating the accuracy of the approximation in terms of the equilibrium geometries. It deserves to be pointed out again that RPA has been shown to give covalent binding energies at least as accurate as GGA for solids \cite{rpa1,rpa3} as well as being highly accurate for weak binding\cite{rpa1,rpa2} and is known to reproduce the appropriate long-range behaviour of the vdW interaction, both from fundamental theoretical considerations\cite{dobson2012,dzyaloshinskii1961} and in its present practical implementation\cite{seb}. As was shown, the equilibrium geometries from RPA are far superior to any other method investigated here, strengthening our confidence in the method. Unfortunately, the comparison of the $C_{33}$ constants has to be seen as inconclusive, both due to the somewhat uncertain experimental data and because of the lack of intralayer relaxations in the RPA calculations. However, we may safely conclude that the $C_{33}$ constants are consistently of the correct magnitude for the materials studied. While we believe that the present method offers benchmark-quality data for vdW bonded systems, it seems to us that further studies using higher order corrections\cite{gruneis2009,hesselmann2011} or approximate kernels in equation (\ref{response})\cite{dobson-wang2000}, are important to put any lingering doubts about the RPA for vdW bonded systems at rest.

\subsection{LDA}
During the course of the present study, it has become clear that calculations using LDA is a very popular choice of method for describing vdW interactions in layered solids. Since the binding properties of the LDA for vdW systems is known to be present by freak occurrence rather than by conscious design, it is of great importance to carefully study the behaviour of the LDA for weakly bonded systems in order to properly characterise the capabilities of this approximation. 
A first drawback of the LDA can be seen when studying the geometries, where the LDA, while often giving a lattice constant that is  close to the experimental number, sometimes gives a very drastic underestimation of the interlayer separation. A large underestimation of the distance is potentially harmful in many applications, such as band gap estimates, since it will exaggerate the influence of the other layers on the electronic structure of a single layer.
For the binding energies, the LDA binding energies show a distribution that has its main weight close to the RPA distribution, but with a split main peak. The compound-by-compound comparison in Figure~\ref{bindencomp} shows, however, that LDA only on average produces binding energies close to the RPA results, and that important features such as the trends in the transition metal dichalcogenides are not reproduced. This shows the that the capability of the LDA to produce correct results for the binding energy of layered materials is limited, and that results based on the LDA should be treated with caution. However, we note that LDA is in fact the functional that to the greatest extent reproduce the experimental trends for the $C_{33}$ constants, although the scattering of the experimental data makes it hard to say how much this should be taken into consideration. It suggests that the behaviour near the equilibrium point is to a large extent determined by the exchange interactions, which are expected to be rather well represented by LDA.
On balance, in view of the consistent performance of LDA over the whole range of layered solids considered here, we must conclude that LDA can be an acceptable approximation, even though we can not recommend it generally. The fact that it is known that the binding comes about for spurious reasons strongly suggests that the LDA should be used only when no other options are available, and restricted to purely descriptive purposes, since the theoretical flaws of the LDA for vdW interaction are so serious that reliable predictions must be considered impossible.

\subsection{DFT-D}
The results for the semi-empirical DFT-D method in comparison with experimental or calculated benchmarks are very scattered. It also shares with LDA the tendency to sometimes produce much too small interlayer separations, but no correlation between the PBE-D and LDA can be seen for the anomalous deviations. This is not overly surprising, as the method is based on a parametrization in terms of entirely atomic quantities, renormalised by fitting to a set of molecules, and we should not expect such a procedure to produce systematically good results for solids. While the extended states of a solid are rather unlike the finite states of a molecule, we still hold it to be plausible that a suitable set of parameters for the DFT-D method can be found on a compound-by-compound basis by fitting of some properties to suitable benchmarks, such as the equilibrium geometry, elastic constants and cohesive energy. That way, the DFT-D procedure, while not suitable for predictive purposes, may still serve as a computationally less demanding option for describing vdW interaction, which is useful in large scale molecular dynamics simulations and similar applications.

\subsection{vdW density functionals}\label{vdw_df_discussion}

The trends among the different vdW density functionals in our comparison are quite clear, with increasing binding energy strength in the sequence vdW-DF1, vdW-DF2, vdW-DF1 (PBE), VV10. Considering the equilibrium geometries as presented in Figures \ref{cdevs}-\ref{thickdevs}, we can see the overall positive trend that the vdW-DF's produce increasingly accurate results as function of the time at which they were published. The original vdW-DF1 gives a too weak vdW binding, which can to some extent be cured by applying the correction on top of the PBE, rather than the RPBE functional. From a principal point of view, this is not an entirely satisfactory solution since it amounts to including a small amount of the spurious LDA exchange binding to correct a too weak dispersion force component in the vdW-DF. It should be noted that the functional that best describes the covalent bonding within the layers is not one of the vdW-DF's, but the PBE functional, of standard GGA type. This underlines the important fact that  the non-local description of correlation must not be viewed as just an addition which sorts out the long-range part of the correlation and that is inert with respect to the covalent interactions.

Our selections of  functionals allows us to make a comparison of the influence of the different ingredients in the functionals. The vdW-DF1 functional (based on RPBE) and vdW-DF1 (PBE) differ only in which GGA approximation that is used for the exchange, yet the results are quite drastically different. This shows that the RPBE functional is overly repulsive, and not the best option for constructing a vdW-DF type functional, and this was one of the original reasons for developing the vdW-DF2 functional\cite{llvdw2}, and to base it on the refitted functional PW86R\cite{murray2009}. Both vdW-DF2 and VV10 are based on the PW86R functional, thus the difference between them is entirely due to the construction of the non-local part of the correlation. We see that the difference is in fact large, the VV10 functional is binding much more strongly than vdW-DF2, as shown both in a good performance for the vdW bond lengths and in a large, but very consistent, overestimate of about 50\% of the binding energy. 
The $C_{33}$ elastic constants for the VV10 functional appear to be somewhat too large in comparison with the experimental data. We believe it safe to conclude that the VV10 functional, while producing excellent geometries, is too stiff and seriously overestimates the binding energy for layered compounds. 

The fact that the present functionals appear to be unable to simultaneously reproduce both good lattice geometries and acceptable binding energies leads to the suspicion that the source of the errors might not be in the non-local correlation. If we revisit our picture of the vdW bonded equilibrium from Section \ref{computational_procedure}, with a strongly repulsive exchange "wall" and a softer vdW "slope", it appears likely that the source of the error should be on the repulsion side of the curve. It is plain to see that the only way to obtain correct interlayer separations with a functional that is overly repulsive in the weakly overlapping regime is to also apply an overly attractive correlation part at large separations. But in doing so one has to pay the price of getting a too large binding energy, since we are overestimating the interaction between the layers. A preliminary investigation for BN showed that a pure Hartree-Fock calculation does indeed give an exchange "wall" significantly shifted towards smaller separations compared with all GGA functionals investigated in the present publication, which suggests that this might be the correct conclusion. This oversimplified comparison is not entirely appropriate since the correlation is included in the GGA calculations. Nevertheless, it is clear that the exchange part of the interaction, as supplied by the underlying GGA functional, is highly important even when determining vdW dominated properties of systems.

\ack
This research was supported by the Academy of Finland through the COMP Centre of Excellence Grant 2006-2011. Computational resources were provided by Finland's IT center for Science (CSC). The ISCD has kindly granted permission to export and store the database in a format appropriate to our purpose.

\appendix
\section{Investigated compounds}
\begin{table}[htdp]
\caption{The compounds investigated in the present study.}
\begin{center}
\begin{tabular}{lc|lc|lc}
1 &   AgBiP$_2$Se$_6$ & 26 & MgI$_2$ & 51 & 1T-TaS$_2$ \\
2 &   BBr$_3$ & 27 &  2H-MoS$_2$ & 52 & 2H-TaS$_2$\\
3 &   BI$_3$ & 28 & 3T-MoS$_2$ & 53 & 1T-TaSe$_2$\\
4 &   BN & 29 & MoSe$_2$  & 54 & 2H-TaSe$_2$\\
5 &   BaFI & 30 & MoTe$_2$ & 55 & 4H-TaSe$_2$\\
6 &   Bi$_2$Se$_3$ & 31 & NbS$_2$ & 56 & Ti$_2$PTe$_2$\\
7 &   Bi$_2$Te$_3$ & 32 & 2H-NbSe$_2$ & 57 & TiS$_2$\\
8 &   BiIO & 33 & 4H-NbSe$_2$ & 58 & TiSe$_2$\\
9 &   C & 34 & NbTe$_2$ & 59 & TiTe$_2$\\
10 &   CdI$_2$ & 35 & Ni$_2$SbTe$_2$ & 60 & TlCrTe$_2$\\
11 &  CoTe$_2$ & 36 & NiSbSi  & 61 & VBr$_2$\\
12 &   CrSe$_2$ & 37 & NiTe$_2$ & 62 & VCl$_2$\\
13 &   CrSiTe$_3$ & 38 & PbBi$_4$Te$_7$ & 63 & VI$_2$\\
14 &   Cu$_2$S & 39 & PbFI & 64 & VS$_2$\\
15 &   Fe(PSe$_3$) & 40 & PbO & 65 & VSe$_2$\\
16 &   GaS & 41 & PbSb$_2$Te$_4$ & 66 & 2H-WS$_2$\\
17 &   GaSe & 42 & PdTe$_2$ & 67 & 3T-WS$_2$\\
18 &   Ge$_2$Sb$_2$Te$_5$ & 43 & PtS$_2$ & 68 & WSe$_2$\\
19 &   HfS$_2$ & 44 & PtSe$_2$ & 69 & Y$_2$I$_2$Ga$_2$\\
20 &    HfSe$_2$ & 45 & PtTe$_2$ & 70 & YI$_3$\\
21 &   HfTe$_2$ & 46 & Re(AgCl$_3$)$_2$ & 71 & ZrNCl\\
22 &   HgI$_2$ & 47 & RhTe$_2$ & 72 & ZrS$_2$\\
23 &   In$_2$Zn$_2$S$_5$ & 48 & SnS$_2$ & 73 & ZrSe$_2$\\
24 &   Mg$_2$(P$_2$Se$_6$) & 49 & SnSe$_2$ & 74 & ZrTe$_2$\\
25 &   MgBr$_2$ & 50 & SrFI & & \\
\end{tabular}
\end{center}
\label{allthecompounds}
\end{table}%

\section*{References}
%\bibliography{vdwrefs2.bib}
%\bibliographystyle{unsrt}

\end{document}